\documentclass[12pt]{iopart}

\newcommand{\J}{J/\psi}
\usepackage{epsfig,epsf}
\usepackage{wrapfig}
\usepackage{graphicx}
\usepackage{epstopdf}
\newcommand{\beq}{\begin{equation}}
\newcommand{\eeq}{\end{equation}}
\newcommand{\lam}{\Lambda_{\rm QCD}}
\def\eq#1{{(\ref{#1})}}

\begin{document}

\title[Theoretical issues in $J/\psi$ suppression]{Theoretical issues in $J/\psi$ suppression}

\author{D E Kharzeev\footnotemark}\footnotetext{Invited plenary talk at the 19th International Conference on Ultra-Relativistic Nucleus--Nucleus Collisions ("Quark Matter"), 14-20 November, 2006, Shanghai, China.}

\address{Nuclear Theory Group,\\ Physics Department, \\ Brookhaven National Laboratory, \\ Upton, NY 11973-5000, USA}\ead{kharzeev@bnl.gov}

\begin{abstract}
Two decades ago Matsui and Satz suggested that Debye screening in the quark-gluon plasma would result  in $\J$ suppression in heavy ion collisions. Much has happened in the subsequent years,  
and the picture of quark-gluon plasma at present is rapidly evolving -- what does it imply for the $\J$ suppression? 
What are the recent RHIC and SPS results trying to tell us? What else has to be done?  
This talk is an attempt to address these questions.
\end{abstract}


\section{Introduction}

This week thirty two years ago the physics world was shaken by what has later become known as "November revolution": the discovery of a new narrow heavy resonance at BNL and SLAC \cite{Aubert:1974js,Augustin:1974xw}. This revolution (unlike many others) is guaranteed to have a lasting effect -- 
first, it added a new "charmed" quark to the list of fundamental building blocks of Nature; second, the explanation of the narrow width of the new resonance  was 
one of the first successes of QCD -- the newly born theory of strong interactions. The large mass of the charmed quark $m_c \simeq 1.3 $ GeV and the smallness of the running QCD coupling constant at this scale $\alpha_s(m_c) \simeq 0.3$  allowed to build a successful theory of "charmonium" decays. Moreover,   
the size of charmonium $R \sim \left(m_c\ \alpha_s(R^{-1})\right)^{-1} \ll \Lambda_{\rm QCD}$ is sufficiently small to justify an attempt of using a perturbative Coulomb potential to describe the bound state structure. 

All of this makes charmonium, and heavy quarkonia in general, an extremely attractive starting point for an 
investigation into the structure of hadrons.  By studying the distortions of the na\"ive "QCD hydrogen atom" picture caused by non--perturbative effects we glean information about the structure of the vacuum. At high temperature, following Matsui and Satz \cite{Matsui:1986dk}, we hope to be able to understand the nature of quark--gluon plasma by 
looking at modifications of the spectral density of heavy quark correlation functions; experimentally, 
we have an access to them by measuring the nuclear modification of charmonium and bottomonium production.

\section{$\J$ in QCD vacuum and the origin of confinement}

Confinement effects in heavy quarkonia are often described by adding a linear term 
to the Coulomb potential, 
\beq\label{cornell} 
V(r) = - {4\over 3} {\alpha_s \over r} + \sigma \ r,
\eeq
where $\sigma$ is the string tension; the resulting "Cornell potential" \cite{Eichten:1978tg} has been shown to describe the spectrum of charmonium states quite well (for a recent review, see \cite{Brambilla:2004wf}). 
Another strategy of accounting for non--perturbative effects is to use the operator product expansion (OPE) and consider the leading power correction to the correlation function of the heavy quark--anti-quark current \cite{Shifman:1978bx} stemming from the dimension four "gluon condensate" $\langle \alpha_s G^2 \rangle$. This procedure is analogous to considering a correction to the energy levels of a hydrogen atom in an external electric field ${\bf E}^2$. Because of the color neutrality of quarkonium, the leading contribution arises from quadratic Stark effect \cite{Voloshin:1978hc,Leutwyler:1980tn}, and the corresponding shift of the energy levels $\delta E \sim \langle \alpha_s\ r^2\ {\bf E}^2 / \epsilon_B \rangle$, 
where $\epsilon_B$ is a typical binding energy. In QCD, the resulting energy shift is positive -- it therefore corresponds to a confining interaction but a one quadratic, not linear, in the inter-quark distance. 

This problem can be seen in the formal OPE expansion for the effective heavy quark potential 
\cite{hq} (see 
\cite{Zakharov03,Dokshitzer:2004ie} for reviews):
\beq\label{ope}
\lim_{r \to 0} V(r) \simeq - \frac{(N_c^2 - 1)}{2 N_c} \frac{\alpha_s(r)}{r} \left(1 + \sum_n a_n \alpha_s^n(r) + c_3 \lam^3 r^3 \right), \label{instpot}  
\eeq
where the second term in the brackets describes the higher order perturbative corrections, and the third one is the 
leading power correction giving rise to the quadratic term in the potential. Since the linear behavior of confining potential has been firmly established by now in lattice calculations, we have to understand the origin of the linear confinement from the point of view of OPE. This is necessary because OPE provides the link to partonic description, and we need it for addressing the problem of quarkonium interactions in matter.  
Several interesting scenarios have been proposed to achieve this goal, and space does not allow me 
to describe them in this talk (see e.g.  \cite{Zakharov03,Dokshitzer:2004ie}); all of them however involve a
modification of the gluon propagator, or the QCD coupling, in the infrared region (the necessity of such a modification is made evident by the Gribov copies problem \cite{Gribov:1977wm} -- the existence of multiple solutions of gauge fixing condition at large coupling). In terms of the OPE these modifications amount to the presence of  {\it dimension two} operator in \eq{ope}; this means that confinement effects manifest themselves already at 
relatively short distances. 

A simple way of realizing such an approach is to introduce a small tachyonic mass into the Coulomb gluon propagator \cite{Chetyrkin:1998yr}: 
\beq\label{tach}
G({\bf q}^2) = {1 \over -\lambda^2 + {\bf q}^2} \simeq {1 \over {\bf q}^2} + {\lambda^2 \over {\bf q}^4},
\eeq
where $\lambda^2 > 0$ is a constant; we assume ${\bf q}^2 \gg \lambda^2 \sim \lam^2$. It is easy to see that a Fourier transform of this propagator $V(r) \sim  \int d^3 q\ \exp(i {\bf q r})\ G({\bf q}^2)$ yields the potential of the form \eq{cornell}.  While this may seem a completely arbitrary procedure, the number of options one can utilize to modify 
the analytical structure of the gluon propagator is quite limited: indeed, confinement means that the gluon propagator cannot have a pole at $q^2 = 0$, or any spectral strength at $q^2 \geq 0$ in general; any modification of spectral density at large $ - q^2 \gg \lam^2$ would spoil a well--established perturbative behavior; and any structure away from the real axis would contradict unitarity. We are thus left only with a possibility of spectral density supported 
at small and negative $q^2 < 0$, and the pole of \eq{tach} is a simple way of achieving this.

\section{$\J$  production at small $x$: strong color field effects}
 
Heavy quarkonium is characterized by three dimensionful scales: heavy quark mass $m_h$, bound state radius $R$ and binding energy $\epsilon_B$ (in perturbation theory, $R \sim (m_h \alpha_s)^{-1}$, $\epsilon_B \sim m_h \alpha_s^2$), which satisfy the hierarchy $m_h \gg R^{-1} \gg \epsilon_B$.  
\begin{wrapfigure}{l}{9cm}
\includegraphics[width=9cm]{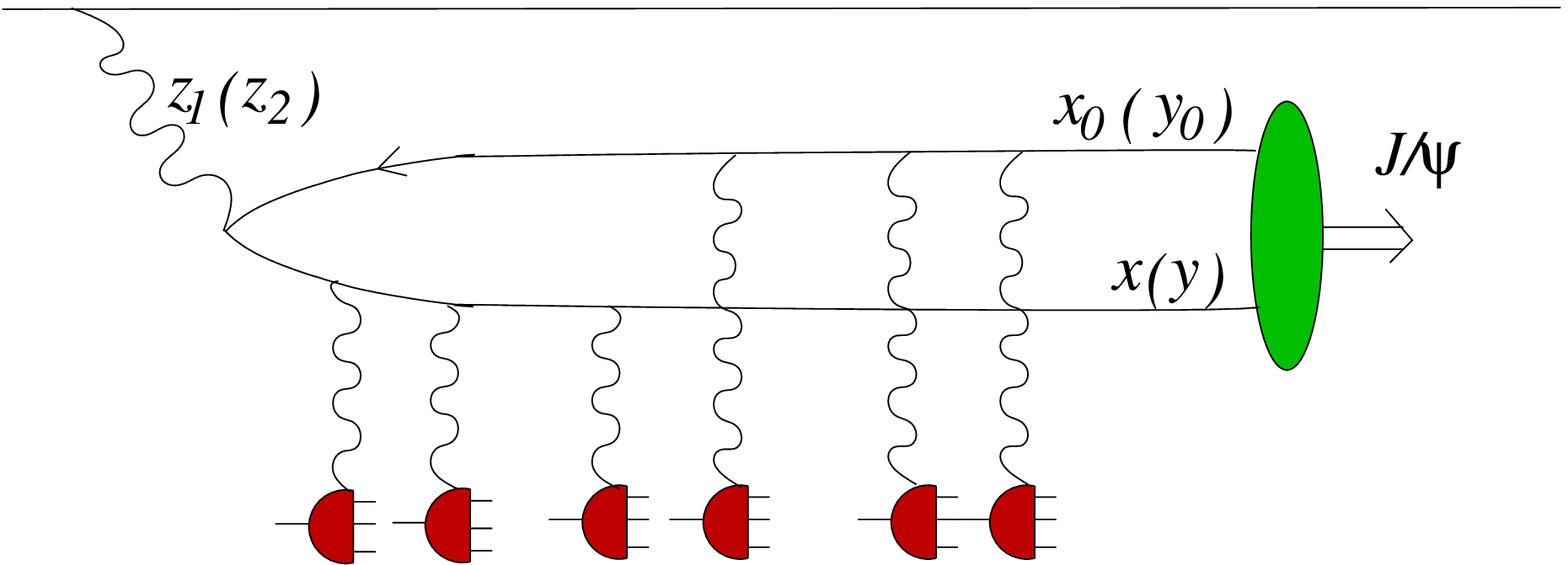} 
\begin{flushleft} 
{\bf Figure 1.} One of the diagrams describing $\J$ production in $pA$ collisions at high energy; from \cite{Kharzeev:2005zr}.
\end{flushleft}
\end{wrapfigure}
The size of light hadrons is $\sim \Lambda_{\rm QCD}^{-1}$, and the  strength of color fields inside them is 
${\bf E}^2 \sim  \Lambda_{\rm QCD}^2$.  Because $m_h \gg \lam$, the production cross section of heavy quark--antiquark pair can be written down in a factorized form as a convolution of partonic sub--process  cross section 
and the parton distributions inside the light hadrons. The corrections to such a factorized form are 
suppressed by powers of $(\lam/m_h)^n \ll 1$.
\begin{figure}
\includegraphics[width=12cm]{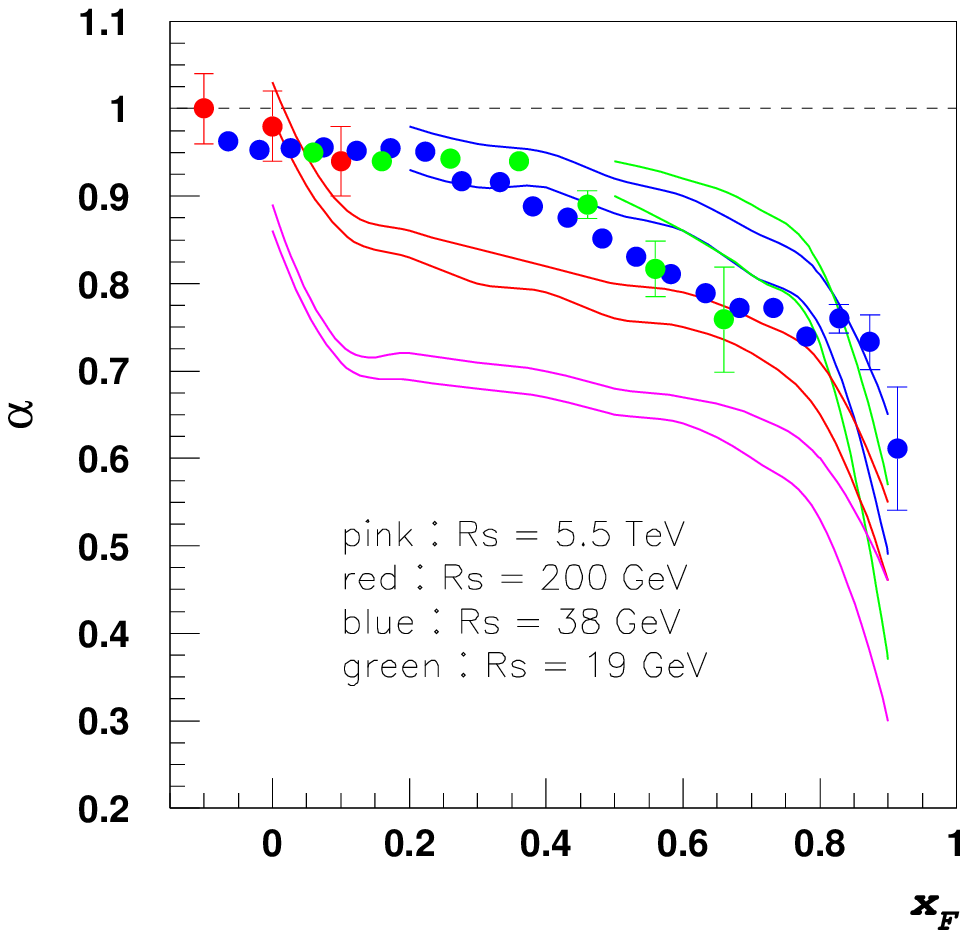} 
\begin{flushleft} 
{\bf Figure 2.} The approximate "$x_F$ scaling" of $\J$ production in $pA$ collisions; from \cite{Kharzeev:2005zr}.
\end{flushleft}
\end{figure}
The situation is much more complicated for the case of quarkonium production, because it happens over 
time scale of $t_{form} \sim \epsilon_B^{-1}$ which is still short, $t_{form} < \lam$ but leaves the possibility 
of significant power corrections -- indeed, the description of $\J$ production in hadron collisions has been 
notoriously difficult.  

These effects are even more pronounced in nuclear collisions and/or at small Bjorken $x$, when the strength of the color field increases significantly and is characterized by the 
saturation scale $Q_s^2(x;A) \gg \lam^2$ increasing with the nuclear mass number $A$ and $1/x$: 
$Q_s^2(x;A) \sim A^{1/3} \ x^{-0.3}$. In this case the powers of $(Q_s/\epsilon_B)^n$ and $(Q_s R)^m$ 
cannot be neglected and have to be resummed. Diagrammatically, these powers correspond to diagrams 
such as the one showed in Fig. 1  \cite{Kharzeev:2005zr}. These diagrams describe the broadening of heavy quark momentum distribution due to multiple scattering inside the target. Because the momentum distribution inside quarkonium is quite narrow (with relative momentum $\sim R^{-1}$) such broadening 
makes it more difficult for heavy quarks to bind, reducing the quarkonium production cross section -- the 
mechanism considered earlier in Ref. \cite{Qiu:1998rz}. The universal behavior of parton densities at small $x$ encoded in the dependence of saturation scale on $x$ translates into an approximate $x_F-$ scaling of $\J$ production in $pA$ collisions \cite{Kharzeev:2005zr}. The increasing shadowing at small $x_2$ (nuclear parton's $x$) is a generic feature of all approaches to nuclear modification of hard processes; however the power corrections to the $\J$ production break down the scaling in $x_2$. The factorization of the nuclear modification factor into a part coming from the nuclear parton distribution and a part describing the final state absorption is severely broken at high energies and/or forward rapidities.
   
\section{$\J$ in the quark-gluon plasma}

Perhaps the most striking theoretical development of recent years is the lattice observation of $\J$ survival 
in quark--gluon plasma at temperatures well above the deconfinement transition temperature $T_c$, up to about $2 T_c$ \cite{Datta:2003ww,Asakawa:2003re}.  This result is in stark contradiction not only with expectations based on perturbative expressions for the screening Debye mass, but also with the calculations of quarkonium dissociation rates 
by thermal gluons 
\cite{Kharzeev:1994pz,Xu:1995eb} (quarkonium break-up in this case is analogous to photo--effect in atoms \cite{Shuryak:1978ij}; this scenario should be relevant at strong coupling, when the binding energy of $\J$ in the plasma is large compared to the temperature \cite{Kharzeev:1996se}).  
\begin{wrapfigure}{l}{9cm}
\includegraphics[width=9cm]{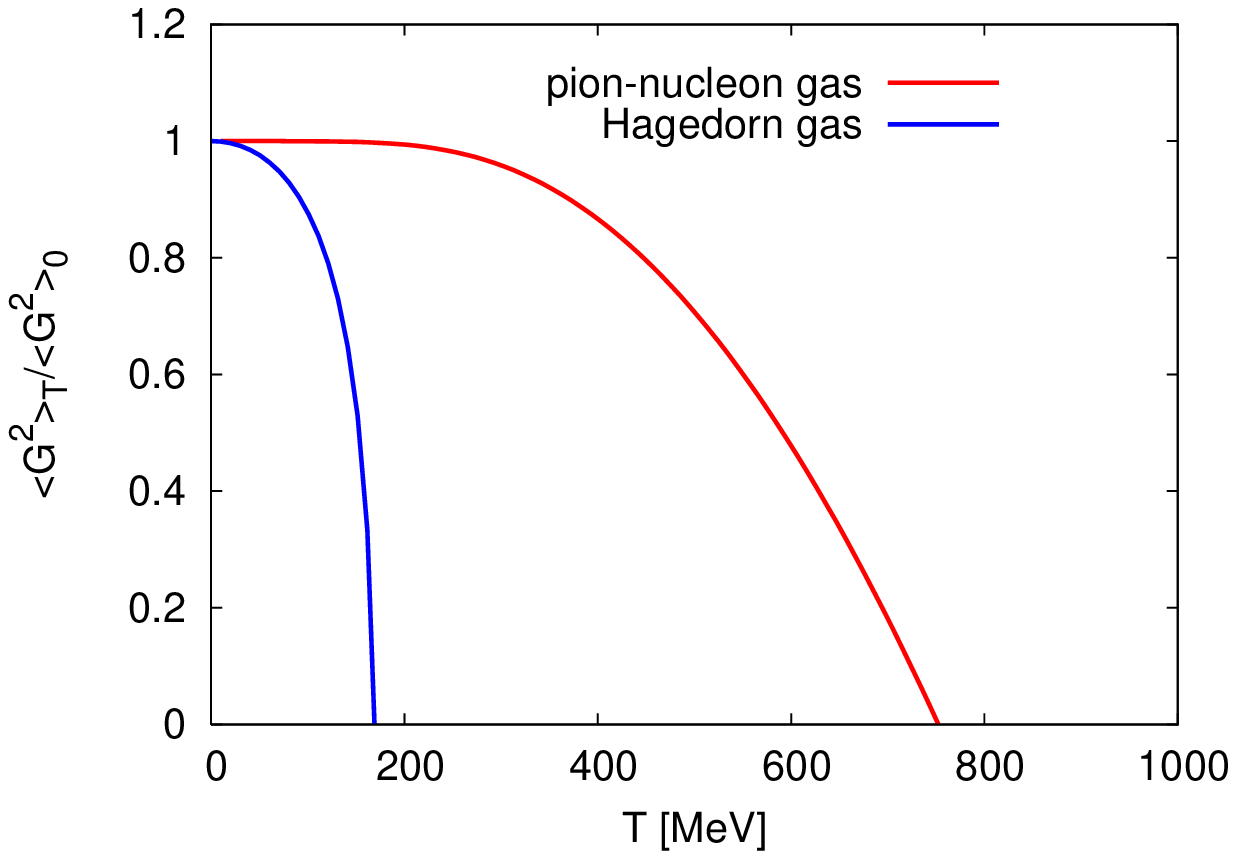} 
\begin{flushleft} 
{\bf Figure 3.} The behavior of gluon condensate as a function of temperature; from \cite{KM}.
\end{flushleft}
\end{wrapfigure}
The basis of these calculations is again the OPE: the dissociation is described by the imaginary part 
of the quarkonium scattering amplitude \cite{Bhanot:1979vb}, whose real part is responsible for the energy shift of quarkonium energy levels as described in the vacuum case by the quadratic piece of the potential \eq{ope}.  In the case of quarkonium in the plasma, the leading operator which is responsible for the mass shift -- the gluon condensate -- becomes temperature--dependent. In hadronic phase, one can evaluate 
this temperature dependence by using the relation 
\beq
\langle \frac{\alpha_s}{\pi}G^2 \rangle_T = \langle\frac{\alpha_s}{\pi}G^2 \rangle_0 + \sum_h g_h\int\frac{d^3p}{2E_\pi(2\pi)^3} n(E_h)\langle h({\bf{p}})|\frac{\alpha_s}{\pi}G^2 |h({\bf{p}})\rangle ,
\label{G2}
\eeq
where $n(E_h)$ are the thermal distributions of hadrons $h$ with degeneracy factors $g_h$;  
the matrix elements of the scalar gluon operator over hadron states can be evaluated with the help 
of scale anomaly. The result \cite{KM} is shown in Fig. 3: even though one can see that pions and even nucleons 
do not affect the condensate significantly below $T_c$, the presence of massive hadron resonances as 
described by the Hagedorn spectrum causes a sharp reduction of the condensate in the 
vicinity of the phase transition. Such a change would certainly affect the mass of $\J$ -- why is it not seen 
in the lattice data? Moreover, lattice results indicate a rapid variation of $\epsilon - 3 p$ above $T_c$; 
since this quantity is related to the gluon condensate, one would expect a change in $\J$ mass also 
in the deconfined phase -- again, a prediction not borne out by the lattice data. What went wrong? 

The puzzle can be solved if the dimension two operator discussed above plays a dominant role in quarkonium dynamics, and if its expectation value does not change significantly at $T \leq 2 T_c$. 
In physical terms, this would mean that residual "short strings" survive well into 
the deconfined phase. To get a rough idea of how this scenario might work, consider the Coulomb propagator \eq{tach} modified by the finite-temperature screening effects \cite{KM}:
\beq\label{tacht}
G(\omega, {\bf q}) = {1 \over - \lambda^2 + \Pi_{00}(\omega, {\bf q}) + {\bf q}^2},
\eeq
where the component $\Pi_{00}$ of the polarization tensor reduces in the static limit to Debye mass squared. 
The Fourier transform of \eq{tacht} yields a potential which exhibits a residual confining interaction even 
above $T_c$, see Fig. 4a). One can define also an effective coupling \cite{Kaczmarek:2004gv} through 
$\alpha(r, T) \equiv {3 \over 4} r^2 {d V(r,T) \over dr}$; the result is shown in Fig. 4b). A comparison of the results of Fig.4 to the 
corresponding lattice results \cite{Kaczmarek:2004gv} reveals a qualitative agreement, surprising in spite of the crude model we used. Recently, thermodynamical properties of the plasma have been investigated in an approach with "Coulomb confinement" in the gluon propagator, with encouraging results \cite{Zwanziger:2004np}.

It is also well known that retardation effects play an important role in quarkonium dynamics; a sharp way to elucidate them is to compare correlation functions evaluated directly on the lattice to the ones computed by using the lattice potential as an input \cite{Mocsy:2004bv} (see also \cite{Wong:2006bx,Alberico:2006vw,Cabrera:2006wh}). Since confinement effects as modeled by the Coulomb propagator 
\eq{tach} are instantaneous and the screening effects are not, we expect that at short time scales the response of quark--gluon plasma will be similar to a response of a confined system even if at large 
time scales the color charges are deconfined. One may speculate that this effect is at the origin 
of both $\J$ survival in the plasma and a strong quenching exhibited by light and heavy partons. 
Regarding $\J$, in this scenario it survives in the plasma despite being bombarded by thermal gluons because the 
heavy quarks are still bound by a confining potential at temperatures $T \leq 2\ T_c$.

\begin{figure}[ht]
  \begin{tabular}{cc} 
      \includegraphics[width=8cm]{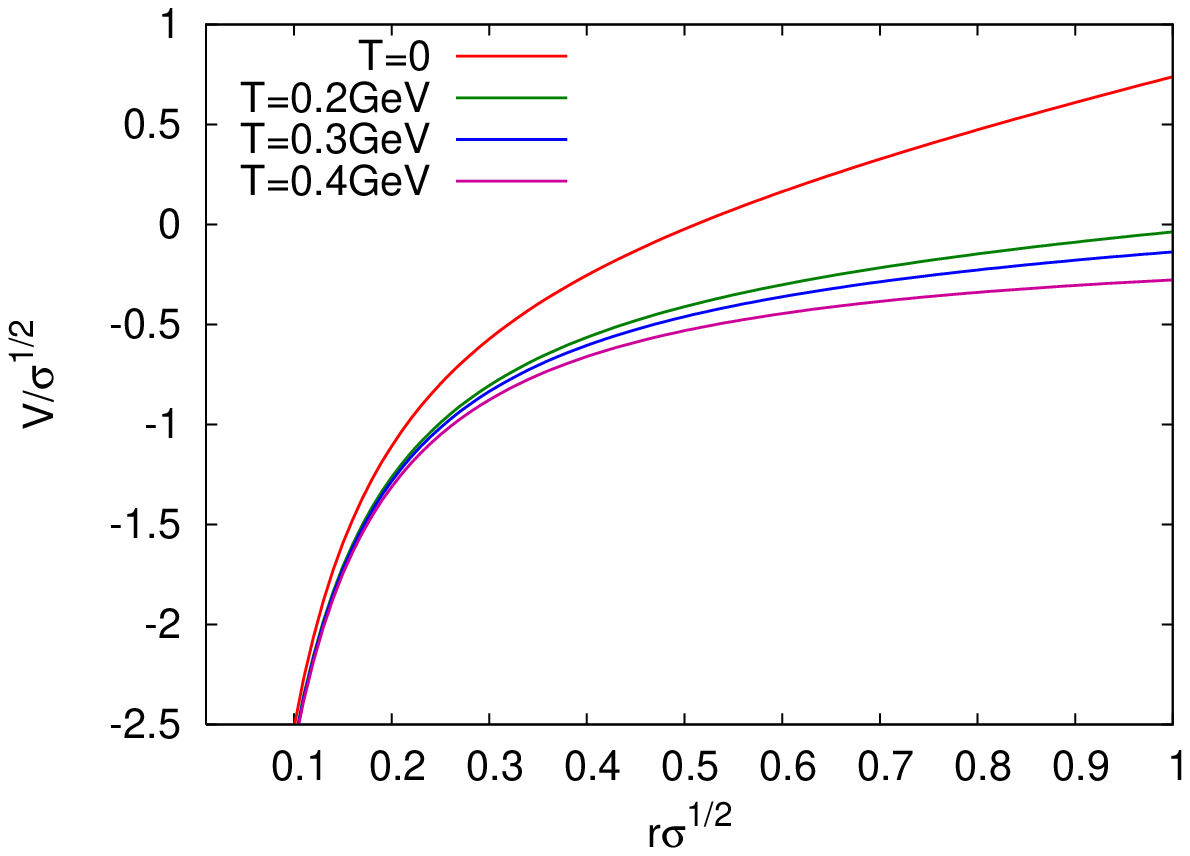} &
        \includegraphics[width=8cm]{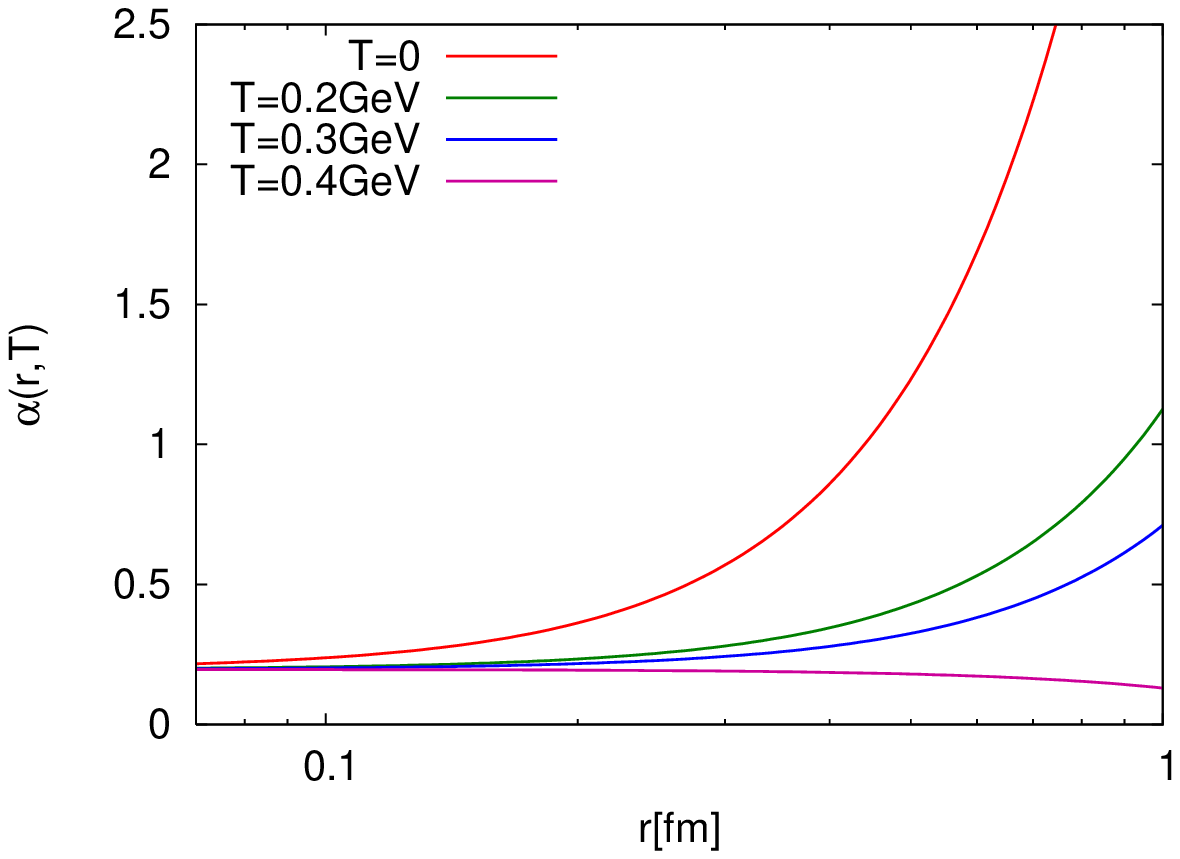}\\
         (a) & (b) 
\end{tabular}
{\bf Figure 4.} a) Static heavy quark potential at different temperatures; b) the corresponding effective strong coupling; from \cite{KM}.
\end{figure}

All of the discussion above assumed that the quarkonium is at rest; how would the picture change for a 
quarkonium traversing the plasma at finite momentum? At weak coupling, the answer has been known 
for quite some time \cite{Chu:1988wh,Mustafa:2004hf}: the gross effect is the decrease of the screening length caused by a larger parton density in the rest frame of heavy quarkonium. Similar result has been 
obtained recently for ${\cal N} = 4$ SUSY Yang--Mills theory \cite{Liu:2006nn} which can be solved even at strong coupling: 
the screening length is reduced by $1/\sqrt{\gamma}$, where $\gamma$ is the Lorentz factor of the moving pair. Experimentally, a stronger screening at finite momentum has to be disentangled from the suppression caused by the gluon fragmentation contribution to $\J$ production expected to dominate at $p_{\rm T} \geq 3 \div 5$ GeV -- since the gluons are known to be quenched, so should the $\J$'s.    

\section{Why are RHIC and SPS results alike?}

High statistics data on $\J$ production in nuclear collisions are available at present from PHENIX \cite{PereiraDaCosta:2005xz,Adare:2006ns,Bickley:2007ac} and NA60, NA50 Collaborations \cite{Arnaldi:2006ee,Arnaldi:2007aa} at RHIC and SPS energies. A surprizing feature of the data is the similarity of nuclear suppression seen at mid-rapidity at RHIC and SPS in spite of a large difference in energy densities achieved at these vastly different energies. A higher density of gluons produced at RHIC was expected to 
\begin{wrapfigure}{l}{9cm}
\includegraphics[width=8cm]{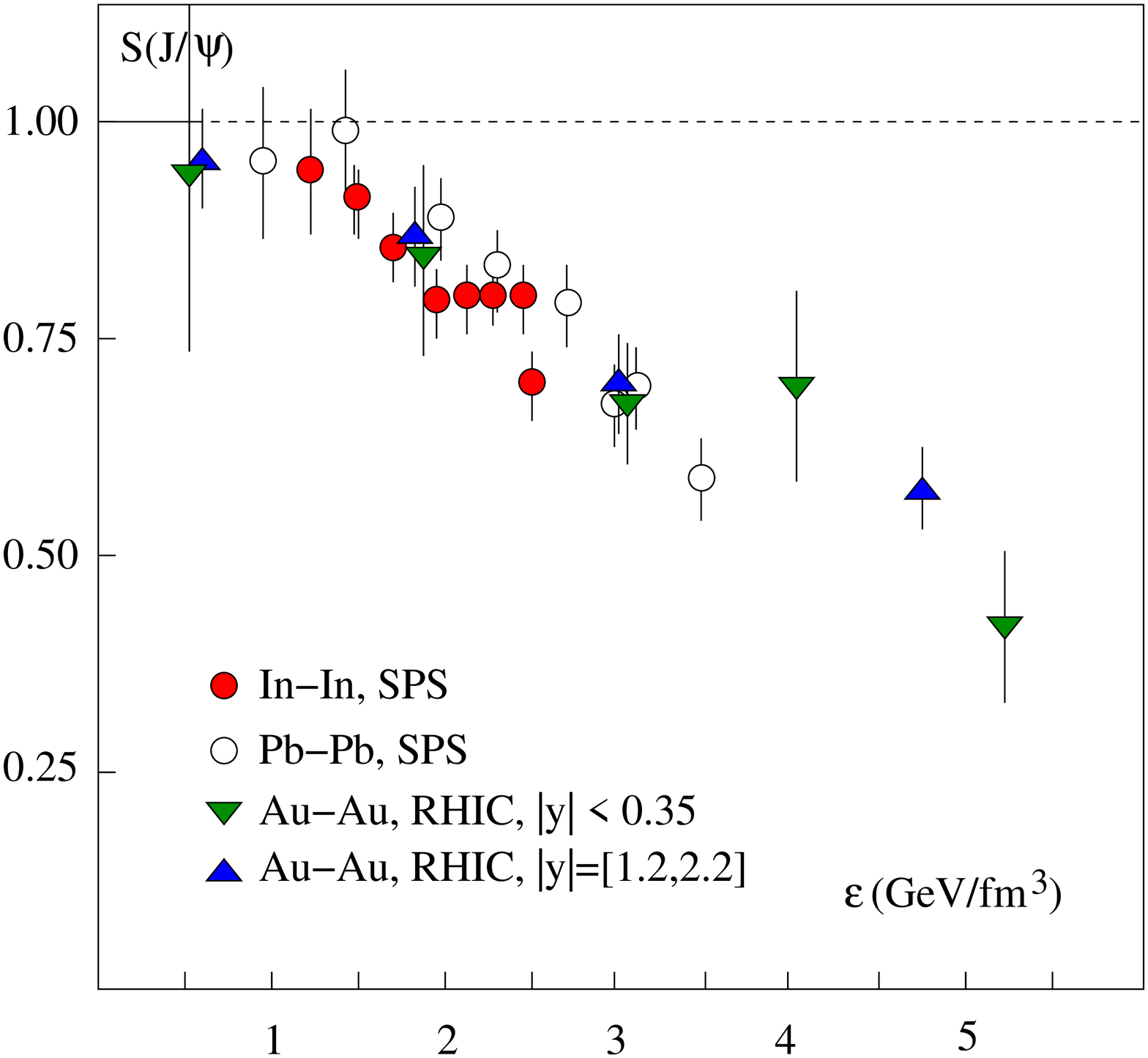} 
\begin{flushleft} 
{\bf Figure 5.} $\J$ suppression at RHIC and SPS; from \cite{Karsch:2005nk}.
\end{flushleft}
\label{supp}
\end{wrapfigure}
result in a stronger suppression -- but what if, as we discussed above, $\J$ survives well into the deconfined phase? The survival of $\J$ up to $\simeq 2\ T_c$ seen in the lattice data means that it can survive up to the energy density $\epsilon \simeq 16\ \epsilon_c$ significantly above the critical one,  
$\epsilon_c \simeq 1\ {\rm GeV/fm}^3$. Given that the {\it initial} energy density in central Au-Au collisions at RHIC is estimated at $\epsilon_{initial} \simeq 20\ {\rm GeV/fm}^3$ (see e.g. \cite{Kharzeev:2000ph,Kharzeev:2001gp}), it is conceivable that even at RHIC the directly produced $\J$'s are still largely intact, and the observed suppression stems from the depletion of $\simeq 40 \%$ of $\J$'s originating from the decays of excited $\chi$ and $\psi'$ states. Such a scenario was put forward in \cite{Karsch:2005nk}; 
it seems to result in a reasonable description of the data, as seen in Fig.5. An alternative explanation invokes the regeneration of $\J$ from charm quark pairs  \cite{Andronic:2003zv,Grandchamp:2003uw,Thews:2005vj}. 
Since the recombination is quadratic in the density of charm quarks, the signature of this mechanism is a narrowing of rapidity and transverse momentum distributions of $\J$'s produced in nuclear collisions. 
As we discussed above, a narrowing of rapidity distribution in nuclear collisions compared to $pp$ ones 
is also expected due to shadowing effects, which suppress $\J$ yields at forward and backward rapidities. 
We now need high statistics $dA$ data to evaluate the contribution of this effect to the observed rapidity dependence of $\J$ suppression in $Au Au$ collisions at RHIC.  Another important measurement is the azimuthal anisotropy of produced $\J$'s.

\vskip0.3cm 

I am indebted to Yuri Dokshitzer, Frithjof Karsch, Eugene Levin, \'Agnes M\'ocsy, Marzia Nardi, Helmut Satz, Kirill Tuchin and Valentin Zakharov for  enjoyable collaborations and stimulating discussions on the topics presented in this talk.  This work was supported by the U.S. Department of Energy under contract No. DE-AC02-98CH10886.

\end{document}